# About set-theoretic properties of one-way functions

Anatoly D. Plotnikov[*]

**Abstract**

We investigate the problem of cryptanalysis as a problem belonging to the class NP. A class of problems $UF$ is defined for which the time constructing any feasible solution is polynomial. The properties of the problems of NP, which may be one-way functions, are established.

## 1 Statement of the problem

There are several equivalent definitions of the problems of the class NP [1, 2]. We say that the problem of $Z$ belongs to the class NP, if:

a) the problem can be determined by a finite number of symbols $n$;

b) the solution of the problem can be represented by a finite the number $m$ of symbols, where $m$ is a polynomial function of $n$: $m = f(n)$;

c) the time of verification of the solution $t$ is a polynomial function of $n$: $t = \varphi(n)$.

It is easy to see that every cryptosystem determines the problems that belong to the class NP.

So, the problem of encryption includes, as an input, $N$-bit plaintext and $M$-bit key, that is, in this case, the problem is determined by $N + M = n$ symbols. The solution of the problem of encryption, as a rule, contains

---

[*]Department of "Computer Systems and Networks" of Dalh East-Ukrainian National University, Luhansk, 91034, Ukraine. E-mail: `a dot plotnikov at list dot ru`

$m = N < n$ bits. Finally, the time of encryption, which can be considered as the time of verification of the obtained solution, is a polynomial function of dimension $n$ of the initial data. Thus, we see that the problem of encryption belongs to the class NP.

The problem of deciphering is symmetric of the problem of encryption. In this case, the initial data are the ciphertext of $N$ bits and $M$-bit key, and the solution of the problem is the $N$-bit plaintext. The time of deciphering is a polynomial function of $N + M = n$. This time of deciphering can also be considered as a verification of the solution. Thus, the problem of deciphering also belongs to the class NP.

There are different types of attacks on a cryptosystem.

Without loss of generality, we can assume that the cryptanalyst, to crack the cryptosystem, has $N$-bit plaintext, $N$-bit ciphertext. He needs to find a solution — the key that contains $M$ bits. In addition, the cryptanalyst knows polynomial-time algorithm for encryption and decryption. In other words, the cryptanalyst has a good algorithm of verification of the received solution. Consequently, the problem of cryptanalysis also belongs to the class NP.

To increase the stability of the cryptosystem, the developers try to make a problem for the cryptanalyst intractable. One way of doing this — to construct one-way function $y = f(x)$, i.e. such that the value of $y$ for the specified $x$ is computed easily (using a polynomial-time algorithm), and inverse operation — computing the inverse function $x = f^{-1}(y)$ — is intractable.

Thus, for example, the process of encryption and decryption consist in to computing some function $y = f(x)$, and the problem of the cryptanalyst consists in computing the difficult inverse function $x = f^{-1}(y)$

It is believed that some functions, using in the existing cryptosystems, are one-way. These are the operations of discrete logarithms, factorization, and others. However, so far no evidence that such functions exist.

The purpose of this paper be to investigate the set-theoretic properties of functions that can be one-way functions.

## 2 The basic constructions

It is clear that the process of constructing solution of the problem of cryptanalysis $Z$ is extensive in the time. That is, theretofore as the solution $Z$ will be found, a cryptanalyst can obtain some intermediate solutions. Denote by $Q$ the set of all intermediate and final solutions of the $Z$. Then each of the final solution $Z$ is called *support solution*, and all solutions of the set $Q$, including support solutions, are *feasible solutions* of $Z$.

Then, more precisely, the problem of the class NP can be formulated as follows:

a) the problem can be determined by a finite number of symbols $n$;

b) any *support solution* can be represented by a finite the number $m$ symbols, where $m$ is a polynomial function of $n$: $m = f(n)$;

c) the time of verification of the *support solution $t$* is a polynomial functions of $n$: $t = \varphi(n)$.

We define a new class of problems $UF$ [3, 5]. We say that the problem of $Z$ belongs to the class $UF$, if:

a) the problem can be determined by a finite number of symbols $n$;

b) any *feasible solution* can be represented by a finite the number $m$ symbols, where $m$ is a polynomial function of $n$: $m = f(n)$;

c) the time of verification of the *feasible solution $t$* is a polynomial functions of $n$: $t = \varphi(n)$.

An example of the problem, in which the finding of an intermediate result requires exponential time of verification of an intermediate solution is the Hamiltonian cycle problem.

**Theorem 1.** $UF \subset NP$ and $UF \neq NP$.

**Proof**. Obviously, that properties of any problem of the class $UF$ are more limited than properties of problem of the class NP because this class does not include such problems of the class NP, for that the construction of some intermediate results requires exponential time, and intermediate results can have exponential length. However, all properties of problem of the class NP are executed also by the problem of the class $UF$. Therefore we have: $UF \subset$ NP and $UF \neq$ NP. Q.E.D.

Problems of the class NP subdivide into "good" and "bad". To "good" we ascribe those problems for which a polynomial-time solution algorithm is known. The set of all such problems forms the class P. Thus, the following correlation is executed: $P \subset NP$.

**Theorem 2.** $P \subset UF$.

**Proof**. Actually, we will suppose that $P \not\subset UF$. Consequently, some intermediate results of calculations of a problem from P are found in the exponential time or have the exponential length. We obtain contradiction to definition of the class P. Q.E.D.

**Corollary 1.** $P \neq NP$.

As it is mentioned in the previous section, the developers of a cryptosystem use the so-called one-way functions. Informally, a one-way function is a function such that it can be easily computed for any input value, but it is difficult to find its argument for the given value of the function. For example, the function $y = b^x$ is an example of such function. Here $b$ is a some number. It is clear that the value $y$ for the given $x$ is computed simply, but the value of $x$ for given $y$ is computed difficulty. Until now, we do not known of efficient (polynomial-time) algorithm to compute the argument of this function.

Let $\Omega$ is the set of all possible one-way functions from the class NP, if they exist.

**Theorem 3.** $\Omega \subset NP \setminus UF$.

**Proof.** Let $\Omega \neq \oslash$ and $y = f(x) \in \Omega$. Clearly, if $y = f(x)$ does not belong to $UF$ then properties b) or c) in the definition of $UF$ will be not executed in process of computing the value of the argument $x$ for a given value $y$. This means that either the length or time of verification of intermediate results will be exponential. As it was required. Q.E.D.

## 3 Conclusion

It needs to make a few comments.

First, it is evidence that the existing definition of the problems of the class NP involves the relation P $\neq$ NP. However, it does not imply the existence of a polynomial-time algorithm for solving all problems of the class $UF$. The question of equality of the classes P $= UF$ are not solved formally. Although the author has proposed a heuristic polynomial-time algorithm to solve one of NP-complete problems from the class $UF$ — the maximum independent set problem, which showed good results [4]. Maybe "worst case" will be found which this algorithm will can not solve. However, until it is not found.

Second, it would be a mistake to believe that on the basis of above-said, the inequality $\Omega \neq \oslash$ is satisfied. As well as to assert that there is no one-way functions. Although it very seems that this is so.

Since the problem of a cryptanalyst belongs to the class NP then any problem of this class is reduced to any NP-complete problem in polynomial time. In particular, to the maximum independent set problem. Therefore, there is a good opportunity to solve this problem in polynomial time.

Therefore, we again arrive at the well-known conclusion: to raise stability of cryptosystems, it is necessary to increase length of the key, not relying on the creation of an "ideal" one-way function, which probably absents in the class NP.